\documentclass[conference]{IEEEtran}
\IEEEoverridecommandlockouts

\usepackage{cite}
\usepackage{amsmath,amssymb,amsfonts}
\usepackage{tikz}
\usetikzlibrary{positioning}
\usetikzlibrary{calc}
\usepackage{array}
\usepackage{xcolor}
\usepackage{listings}
\usepackage{colortbl} 
\usepackage{graphicx}
\usepackage{subcaption}
\usepackage{makecell} 
\usepackage{enumitem}
\usepackage{amsthm}
\usepackage{amsmath}
\usepackage{algorithm}
\usepackage{algorithmic}
\usepackage{multirow} 
\usepackage{algorithmic}
\usepackage{graphicx}
\usepackage{textcomp}
\usepackage{xcolor}
\usepackage{url}
\def\BibTeX{{\rm B\kern-.05em{\sc i\kern-.025em b}\kern-.08em
    T\kern-.1667em\lower.7ex\hbox{E}\kern-.125emX}}
\begin{document}

\newtheorem{definition}{Definition}

\title{Data Guard: A Fine-grained Purpose-based Access Control System for Large Data Warehouses\\
}

\author{
\IEEEauthorblockN{Khai Tran\textsuperscript{*}, Sudarshan Vasudevan\textsuperscript{*}, Pratham Desai\textsuperscript{*}, Alex Gorelik\textsuperscript{}, Mayank Ahuja\textsuperscript{}, \\
Athrey Yadatore Venkateshababu\textsuperscript{}, Mohit Verma\textsuperscript{}, Dichao Hu\textsuperscript{}, \\
Walaa Eldin Moustafa\textsuperscript{}, Vasanth Rajamani\textsuperscript{$\ddagger$}, Ankit Gupta\textsuperscript{$\S$}, Issac Buenrostro\textsuperscript{}, Kalinda Raina\textsuperscript{}}
\thanks{\textsuperscript{*}Equal Contributors}\thanks{\textsuperscript{$\ddagger$}\textsuperscript{$\S$}Work done while at LinkedIn Corporation}
\IEEEauthorblockA{
\textsuperscript{}\textit{LinkedIn Corporation}, Sunnyvale, CA\\
\textsuperscript{$\ddagger$}\textit{OpenAI}, San Francisco, CA\\
\textsuperscript{$\S$}\textit{Anthropic}, San Francisco, CA\\
\{khtran, suvasudevan, padesai, agorelik, mahuja, avenkateshababu, moverma, dihu,\\
wmoustafa, ibuenros, kraina\}@linkedin.com, vasanth@openai.com, ankit@anthropic.com}
}

\maketitle

\begin{abstract}
The last few years have witnessed a spate of data protection regulations in conjunction with an ever-growing appetite for data usage in large businesses, 
which presents significant challenges for businesses to maintain compliance. 
To address this conflict, we present \emph{Data Guard} - a fine-grained, purpose-based access control system for large data warehouses. 
Data Guard enables authoring policies based on semantic descriptions of data and  purpose of data access. Data Guard then translates these policies into SQL views that mask data from the underlying warehouse tables. 
At access time, Data Guard ensures compliance by transparently routing each table access to the appropriate data-masking view based on the purpose of the access,  
thus minimizing the effort of adopting Data Guard in existing applications.
Our enforcement solution allows masking data at much finer granularities than what traditional solutions allow. 
In addition to row and column level data masking, Data Guard can mask data at the sub-cell level for columns with non-atomic data types such as structs, arrays, and maps. 
This fine-grained masking allows Data Guard to preserve data utility for consumers while ensuring compliance. We implemented a number of performance optimizations to minimize the overhead of data masking operations. 
We perform numerous experiments to identify the key factors that influence the data masking overhead and demonstrate the efficiency of our implementation. 
Data Guard is deployed inside LinkedIn's production data warehouses and ensures compliance of more than 20,000 table accesses each day across different data processing engines.  
\end{abstract}


\vspace{-1em}
\section{Introduction} \label{intro}
The last two decades have witnessed a tremendous growth both in the collection and usage of data to power use cases such as business analytics, recommender systems and personalization. A significant amount of data that businesses collect from end users to power these applications, is sensitive or personally identifiable. Understandably, there have been concerns from regulators, privacy advocates and various other citizen groups alike concerning the lack of transparency in the data collection and data handling practices adopted by businesses. Regulations such as GDPR \cite{GDPR16}, CCPA \cite{CCPA18}, HIPAA \cite{HIPAA03}, DMA \cite{DMA24} and EU AI Act~\cite{AIAct24} have been enacted into law in response to these privacy concerns and to drive transparency into the data collection and data usage practices of businesses. Specifically, these regulations require businesses to limit the usage of user data only for the purposes for which it is collected. As an example, the GDPR regulation requires businesses to seek consent from their end users (or formally, data subjects) in order to use sensitive attributes such as age and gender for advertising purposes.  At the same time, the use of this data is often permitted in anonymized form to enable business use cases intended for demographic analysis. 

To address these privacy concerns and ensure compliance with regulations, several solutions have been proposed~\cite{Byun05, Colombo17, Xue23} to enforce access control policies based on the purpose of data access. The systems are commonly referred to as \emph{Purpose-based Access Control} (PBAC) systems. Depending on the granularity of enforcement, these systems can either authorize/deny access to an entire dataset (\emph{coarse-grained} access control), or they can apply dynamic masking to hide contents of individual rows, columns or cells within the dataset (\emph{fine-grained} access control).  Enforcing policies dictated by these regulations across a large organization can be a challenging endeavor for the following reasons: 
\begin{enumerate}
\item A typical modern data stack involves multiple data processing systems such as Apache Spark~\cite{Spark}, Trino~\cite{Trino} and Apache Flink~\cite{Flink}. Thus, any access control solution is required to apply data access policies consistently across a diversity of data processing systems used in an organization.
\item Many data-centric organizations commonly support 10-100s of thousands of datasets, 10-100s of use cases (or {\em purposes}) and 100s-1000s of data pipelines accessing exabyte-scale data every day. A practical solution for access control must therefore be sufficiently scalable to meet these demands.
\item The data models underlying these datasets can be varied and complex. For instance, it is common to model data using non-atomic data types like arrays, maps, and structs. Furthermore, it is also common for AI/ML workloads to employ application-specific descriptions of data overlaid on basic data structures such as arrays and structs. A practical solution for access control must support granular data-masking capabilities (at a {\em sub-cell} level) within such complex data structures to maximize data utility while ensuring compliance. 
\end{enumerate} 

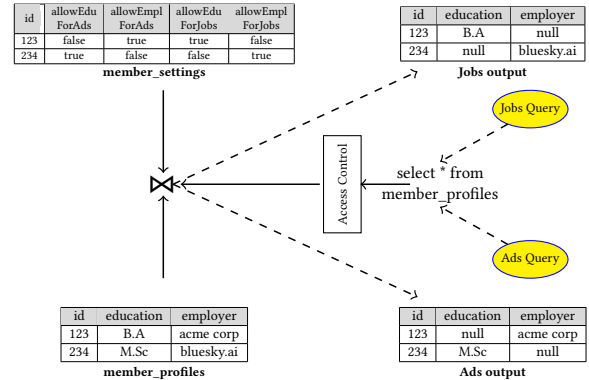
\begin{figure}[ht]  
\begin{tikzpicture}
    \coordinate (left) at (0,-2); 
    \coordinate (left_top) at (0,2); 
    \coordinate (right_top) at (4.5,2);  
    \coordinate (right_bottom) at (4.5,-2); 
    \coordinate (mid_text) at (3.8,0); 
    \coordinate (jobs) at (5,1);
    \coordinate (ads) at (5,-1);
    \coordinate (rect) at (2.5,0);
    \coordinate (join) at (0.1,0);
    
    \def\scalefactor{0.5} 
    \def\scalefactorlefttop{0.5} 
    \def\scalefactorfaces{0.1} 

    \node at (left_top) {
        \scalebox{\scalefactorlefttop}{
  	   \begin{tabular}{|c|c|c|c|c|}
            \hline
            \rowcolor{gray!30} \makecell{id} & \makecell{allowEdu\\ForAds} & \makecell{allowEmp\\ForAds} & \makecell {allowEdu\\ForJobs} & \makecell{allowEmp\\ForJobs} \\
            \hline
            123 & false & true & true & false \\
            \hline
            234 & true & false & false & true \\
            \hline
        \end{tabular}
        }
    };
    \node[below=1em of left_top] {\tiny{\textbf{member\_settings}}};

    \node at (left) {
        \scalebox{\scalefactor}{
        \begin{tabular}{|c|c|c|}
            \hline
            \rowcolor{gray!30} id & education & employer \\
            \hline
            123 & B.A & acme corp \\
            \hline
             234 & M.Sc & bluesky.ai \\
            \hline
        \end{tabular}
        }
    };
    
    \node[below=1em of left] {\tiny{\textbf{member\_profiles}}};

    \node at (right_top) {
        \scalebox{\scalefactor}{
        \begin{tabular}{|c|c|c|}
            \hline
            \rowcolor{gray!30} id & education & employer \\
            \hline
            123 & B.A & null \\
            \hline
            234 & null & bluesky.ai \\
            \hline
        \end{tabular}
        }
    };
    \node[below=1em of right_top] {\tiny{\textbf{Jobs output}}};

    \node at (right_bottom) {
        \scalebox{\scalefactor}{
        \begin{tabular}{|c|c|c|}
            \hline
            \rowcolor{gray!30} id & education & employer \\
            \hline
            123 & null & acme corp \\
            \hline
            234 & M.Sc & null \\
            \hline
        \end{tabular}
        }
    };
    \node[below=1em of right_bottom] {\tiny{\textbf{Ads output}}};
    
    \node at (mid_text) {
        \scalebox{0.8}{
             \begin{tikzpicture}
                \node at (2,0.75) {\small\makecell{select * from \\ member\_profiles}}; 
            \end{tikzpicture}
        }
    };

    \draw[draw=blue, fill=yellow] (jobs) ellipse (0.5cm and 0.25cm);
    \node at (jobs) {\tiny{Jobs Query}};

    \draw[draw=blue, fill=yellow] (ads) ellipse (0.5cm and 0.25cm);
    \node at (ads) {\tiny{Ads Query}};

    \draw[->,dashed,line width=0.2mm] (4.7,0.8) -- (3.8,0.30);
    \draw[->,dashed,line width=0.2mm]  (4.7,-0.8) -- (3.8,-0.30);

    \node[draw, rectangle, minimum width=0.5cm, minimum height=1cm] (rect) at (rect) {\rotatebox{90}{\tiny{Access Control}}};
     \draw[->,line width=0.2mm]  (3.4,0) -- (2.74,0);

    \node at (join) {\huge$\Join$};
     \draw[->,line width=0.2mm]  (2.2,0) -- (0.35,0);
     
     \draw[->,line width=0.2mm]  (0.12,1.25) -- (0.12,0.15);
     \draw[->,line width=0.2mm]  (0.12,-1.25) -- (0.12,-0.15);

     \draw[->,dashed,line width=0.2mm]  (0.25,0) -- (3.5,1.5);
     \draw[->,dashed,line width=0.2mm]  (0.25,0) -- (3.5,-1.5);

\end{tikzpicture}
\caption{An example highlighting dynamic masking of data to honor member preferences based on purpose yielding different outputs.}  
\label{fig:Data Filtering}  
\end{figure}

There are additional challenges for data-centric applications (such as LinkedIn among others) that employ a consent framework that controls how end-user data is used for different use cases. For instance, LinkedIn provides granular consent settings to its over one billion users (or {\em data subjects}) giving them control on how their data is used by LinkedIn for different use cases. 
Thus, data present in warehouse tables need to be dynamically masked based on data subject preferences in addition to the purpose of access. 
In Figure~\ref{fig:Data Filtering}, we illustrate how access control based on data subject preferences differs from the more common row and column-level access control used in practice. 
In this example, the {\em member\_profiles} table storing data subject information is queried for two different use cases, viz. “Ads” and “Jobs”. 
Data subject preferences controlling usage of different attributes of their data are stored in the \emph{member\_settings} table. 
Usage of data subject information for the Ads use case is controlled by two settings, \emph{allowEduForAds} and \emph{allowEmpForAds}, 
while that for Jobs use case is controlled by the settings \emph{allowEduForJobs} and \emph{allowEmpForJobs}. 
Based on the setting values for \emph{allowEduForAds} and \emph{allowEmpForAds}, 
the education attribute associated with $memberId=123$ and employer attribute corresponding to $memberId=234$ in the {\em member\_profiles} table need to be masked for the Ads use case. 
Similarly, the education attribute of member $234$ and the employer attribute of members $123$ need to be masked for Jobs use case based on the corresponding values for jobs-related settings.
\vspace{-0.8em}
\vspace{-1em}

Existing solutions have limited fine-grained access control capabilities and fall short in addressing one or more of the aforementioned challenges. Furthermore, we are not aware of any prior literature describing the myriad practical challenges encountered when deploying an access-control solution across a large data warehouse supporting a diversity of data processing systems.
We therefore implemented Data Guard, a fine-grained purpose-based access control system for large data warehouses. The system translates policies specified in a domain-specific language (DSL) into purpose-specific views that mask data at access time. This translation is facilitated by maintaining a taxonomy of data classification labels (henceforth, referred to as \emph{policy labels}), which are used to provide semantic description of the data elements (e.g. rows, columns, and cells in a table) present in the warehouse. The policies use purpose and policy labels to define the conditions under which access to a data element described using an annotation is allowed or denied.  A \emph{policy compiler} translates these rules into a purpose-specific view SQL and registered with the warehouse. At access time, a view routing layer transparently routes accesses to the appropriate data-masking view based on the purpose attached to the access.  

The key contributions of this paper are as follows:
\begin{itemize}
    \item Data Guard has been successfully deployed across LinkedIn's exabyte-scale data warehouse and ensures that data accesses using Spark and Trino -- 
    the data processing engines used to query the warehouse data at LinkedIn -- are compliant with our data usage policies, and honor preferences of $>$1 billion LinkedIn members.
\item The implementation of Data Guard involves a number of novel ideas that ensure both compliance and ease of use for warehouse users and application developers such as: 
    (i) High-level abstractions that allow policies to be defined using purposes and semantic data labels. These policies are translated by Data Guard into data-masking views that serve as the data access API. 
    (ii) A routing layer referred to as {\em ViewShift} which transparently routes accesses to warehouse tables from user queries to the appropriate data-masking views provisioned by Data Guard. 
    (iii) View optimizations to minimize performance overhead in existing data processing jobs, such as the use of memory-efficient bitmaps in lieu of expensive table joins.  
    (iv) Use of engine-agnostic views which allow data-masking views to be executable across different data processing engines (Spark and Trino at LinkedIn).
 \item To the best of our knowledge, Data Guard is the first practical implementation of an access control system that allows data masking at the sub-cell level. 
    Data Guard achieves this via an expressive grammar that allows it to locate and mask deeply nested data elements inside composite attributes. 
    These granular masking capabilities of Data Guard allows data consumers to maximize value from the underlying data while still ensuring compliance. 
\end{itemize}  

At LinkedIn, Data Guard is currently being used to enforce more than 110 unique data access policies for more than 50 purposes across two major production warehouses. Across the two major production warehouses, Data Guard actively maintains $\sim$13K views on $\sim$4K tables, having created more than 270K versioned views overall as labels and policies are updated. Each day, more than 20K table accesses across Spark and Trino are protected by Data Guard to ensure they are compliant with the defined data policies. 

In the rest of the paper, we provide a detailed description of the design of Data Guard, the key architectural choices in our system along with the reasons that motivated these choices, and the system implementation details. 
We also present data that quantifies the performance overhead of enforcement during data access and conclude with interesting areas of future work.
\vspace{-0.8em}

\section{Related Work}
Fine-grained access control in database management systems (DBMS) has been extensively studied, with numerous mechanisms proposed over the decades~\cite{Stonebraker74, Rizvi04,Xue23, Agrawal05, Pappachan20} that attempt to modify user's data access queries into compliant ones. Early approaches such as~\cite{Stonebraker74}, dynamically modified user queries to enforce attribute-level restrictions. More recently, Xue et. al.~\cite{Xue23}, propose a similar solution specifically for Spark applications by introducing an access control enforcement stage in Spark's Catalyst optimizer. ~\cite{Rizvi04} employs statically defined authorization views which encode data access policies and admits a user query only if it can be mapped to an existing authorization view.~\cite{Pappachan20} enforces policies by dynamically rewriting queries to efficiently apply row-level masking. While effective in their original settings, these solutions fall short in modern data stacks, where access control often requires sub-cell-level masking and enforcement across heterogeneous processing engines.

Another line of work targets policy violation detection via static program analysis~\cite{Wang221,Wang222,Sen14}. These approaches verify compliance but do not provide compliant query results at runtime. By contrast, Data Guard’s contribution is to automatically return compliant data to consumers, achieved through dynamic masking that incorporates both organizational policies and data-subject preferences.

The design in Agrawal et al.~\cite{Agrawal05} shares several ideas with Data Guard, such as semantic descriptions of data and preference-aware filtering of tuples and attributes. However, their system relies on dynamically generated views to enforce restrictions. Further, the solution in~\cite{Agrawal05} is described in the context of relational databases and does not therefore consider non-relational data models and sub-cell masking. In contrast, Data Guard employs static versioned views, a key technical choice that ensures query reproducibility and debuggability, both critical for production-scale analytics. Further, Data Guard is designed for modern data warehouses that allow non-relational data models and require sub-cell-level masking.

Commercial systems such as Snowflake~\cite{Snowflake}, Databricks~\cite{Databricks}, and Amazon Redshift~\cite{Redshift} support dynamic masking via attribute-level semantic tags. Yet, they inherit the drawbacks of~\cite{Agrawal05}: lack of reproducibility, limited debuggability, and insufficient scalability for policies requiring large joins (e.g., user consent tables). Moreover, these systems generally cannot support sub-cell-level masking, which Data Guard introduces to maximize data utility while preserving compliance.

Purpose-based access control mechanisms proposed in~\cite{Byun05, Bertino05, Byun08} enrich relational models by associating intended usage purposes with data objects and filtering queries based on purpose compatibility. While conceptually elegant, these solutions are restricted to tuple-level enforcement, lack support for columnar or sub-cell masking, and require data rewriting when purposes evolve—which is impractical in large dynamic warehouses. Data Guard avoids such limitations by decoupling enforcement from data rewriting and supporting multiple levels of masking granularity.

Policy specification languages have also been explored in depth~\cite{Cranor02,Ni08,Xacml3,Sen14}. Data Guard’s policy language draws inspiration from P3P~\cite{Cranor02} in its use of semantic tags to describe protected data. However, unlike these earlier efforts, we found no off-the-shelf compilers capable of translating such high-level specifications into enforcement-ready representations. Consequently, we designed a lightweight, constrained policy language and custom parser tailored for our current needs.

Finally, privacy-preserving databases~\cite{Agrawal02, Chawla05, Dwork06} provide orthogonal but complementary mechanisms. For instance, Agrawal et al.~\cite{Agrawal02} presented an early privacy-preserving DBMS architecture, while Dwork~\cite{Dwork06} introduced differential privacy, and Chawla et al.~\cite{Chawla05} formalized privacy metrics. While these works do not directly address access control, Data Guard’s flexible enforcement framework can incorporate such privacy-preserving techniques, extending its applicability to privacy-constrained analytics.
\section{Architecture Overview} \label{sec:overview}
\begin{figure*}[ht]  
    \centering 
    \vspace{-1.6em}
    \includegraphics[width=0.7\textwidth]{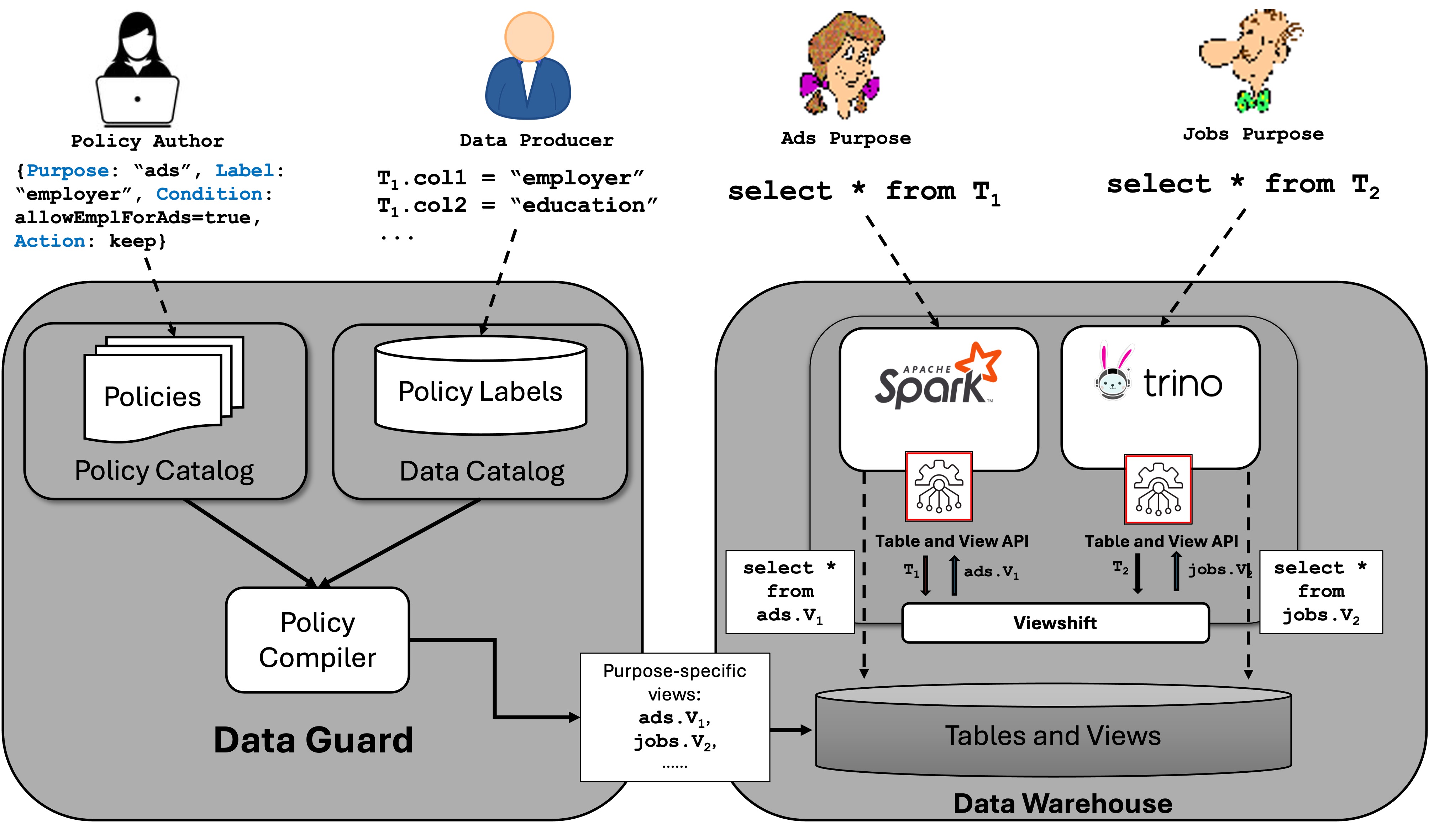} 
    \caption{Data Guard System Architecture.}  
    \label{fig:SystemOverview}  
    \vspace{-1.6em}
\end{figure*}
Figure~\ref{fig:SystemOverview} shows the overall system architecture of Data Guard. 
Data producers provide semantic descriptions of data in the form of {\em policy labels} that are assigned to {\em fields} in a table. 
{\em Fields} refer to data elements in a table and represent the unit at which data masking is applied. 
Thus, a field can reference a column, a row, a cell or portion of data within a cell. 
The mapping of policy labels to fields is stored in a {\em Data Catalog} as shown in Figure~\ref{fig:SystemOverview}. 
Policies in Data Guard are authored in a domain-specific language (DSL) and stored in a {\em Policy Catalog} which provides APIs to query policies from the catalog. 
Each policy is assigned to a purpose and policy label pair. A policy definition has an associated boolean condition that determines whether a data element with the corresponding policy label is masked or preserved during data access.       
Figure~\ref{fig:SystemOverview} shows an example policy that is assigned to $ads$ purpose and $employer$ label. 
The {\em Policy Compiler} consumes the policy label-field mapping and policy definitions as inputs and generates purpose-specific static views which are registered with the data warehouse. 
User queries submitted via Spark or Trino that access tables (e.g. $T_1$ and $T_2$ in Figure~\ref{fig:SystemOverview}) are transparently redirected by a routing module called {\em ViewShift} to the appropriate data-masking views ($jobs.V_1$ and $ads.V_2$ respectively in Figure~\ref{fig:SystemOverview}) based on the accessor's purpose (i.e $jobs$ and $ads$ in Figure~\ref{fig:SystemOverview}). 
We next describe the access control model and system implementation details of Data Guard architecture.

\section{Access Control Model}
In this section, we introduce the foundational abstractions of Data Guard: (i) data access policies, (ii) data-masking views, and (iii) field paths. 
Together, these concepts enable the construction and deployment of a scalable access control system at LinkedIn that supports both fine-grained data protection and purpose limitation.

\subsection{Data Access Policies}
A data access policy in Data Guard specifies: (i) a {\em purpose} under which the policy is applicable, (ii) a {\em label} that determines the data elements to which the policy applies, and (iii) a set of access control {\em rules} that specify the conditions under which access to a data element with a given label is allowed or denied. We next introduce each of these concepts followed by a formal definition of a data access policy. 
 
\subsubsection{Purpose}
The access control model described in this paper is motivated by the {\em purpose limitation} principle introduced in the GDPR regulation~\cite{GDPR16}. Purpose limitation requires that data is collected from {\em data subjects} (i.e. individuals whose personal data is collected) for an explicitly declared purpose and is only processed for use cases compatible with the purpose under which it is collected. Further, usage of an individual's data for newer purposes is driven based on consents from the individual. At LinkedIn, a purpose is used to indicate the business justification for data collection or data processing. Purposes can either map to an external product such as advertising and jobs, or correspond to internal use cases such as security and business analytics. 

Purpose limitation is ensured by requiring each data access to specify a valid purpose. At LinkedIn, all production workloads access warehouse data using service accounts. Similar to role-based access control systems where roles are assigned to a user, we assign a purpose to each service account. The purpose of the service account determines the access control policies that are enforced by Data Guard on the accessed tables. 

\subsubsection{Policy Labels}
Data producers provide semantic description of data in the tables they produce using {\em policy labels}. Figure~\ref{fig:SystemOverview} shows a data producer assigning policy labels $employer$ and $education$ assigned to $col_1$ and $col_2$ of table $T_1$. In general, data producers can assign policy labels either to:
(i) a table, (ii) column in a table (either a top-level or a nested column), (iii) cells, or (iv) sub-cell-level data elements within arrays, structs and maps. This fine-grained assignment of policy labels allows Data Guard to apply data masking at much finer granularities than existing solutions. 

Figure~\ref{fig:SystemOverview} also shows that policy labels are stored as metadata inside a {\em Data Catalog}. This design choice is motivated by recent trends in data architecture where data catalogs such as Horizon~\cite{Horizon}, Unity~\cite{Unity} and DataHub~\cite{Datahub} have emerged as the foundational infrastructure component to address data discovery and data governance needs of large organizations. These catalogs serve as the inventory of the critical data assets (e.g. warehouse tables) of an organization and store a variety of metadata such as schemas, lineages and ownership associated with these assets. These catalogs are therefore well-suited for storing compliance metadata such as {\em policy labels}. 

Policy labels link policies to the data elements (e.g. tables, columns, and cells) to which policies apply. The cardinality of policy labels is typically much smaller than the number of data assets and fields across all tables in a warehouse. Defining policies using policy labels significantly reduces policy duplication and allows policies to be consistently applied across tables in a warehouse as well as across non-warehouse data assets such as online data stores and data streams. Further, it automatically ensures compliance of newly created tables containing data elements tagged with one or more previously defined labels. 

\subsubsection{Policy Language} \label{subsubsec:policies}
Data masking in Data Guard is controlled using rules that are assigned to a purpose and policy label pair. 
\begin{definition} \label{def:policy}
Let $\mathcal{G}$ denote the set of purposes and $\mathcal{L}$ denote the set of policy labels. A policy $p$ in Data Guard is a tuple $(g, l, \langle c, a \rangle)$ where $g \in \mathcal{G}$ and $l \in \mathcal{L}$. $\langle c, a \rangle$ is a condition and action pair (or, a rule) where:
\begin{itemize}
\item $c$ is a compound SQL predicate composed of simple predicates connected using the boolean operators $\{\texttt{AND}, \texttt{OR}, \texttt{NOT}\}$. Each simple predicate in $c$ is of the form $(x \mathbin{\circ} y)$ where:
\begin{enumerate}
\item $x \in \mathcal{X}$ is a set of attributes (defined below),
\item $y \in \mathrm{dom}(x)$ is the domain of values of attribute $x$.
\item $\circ$ represents a relational operator from the set of relational operators $\{=, \neq, <, >, \leq, \geq, \texttt{BETWEEN}, \texttt{IN}, \texttt{LIKE}, \texttt{IS NULL}\}$.
\end{enumerate}
\item $a \in \mathbb{A}$ represents an action from the set of actions $\mathbb{A}$ that the access control system must take when $c$ is true.
\end{itemize} 
\end{definition}
Thus, the example policy shown in Figure~\ref{fig:SystemOverview} can be represented as a tuple:
\vspace{-0.5em}
\begin{equation} \label{eq:policyex}
(ads, employer, \langle allowEmpForAds=true, \mathrm{KEEP} \rangle)
\end{equation}

The set $\mathcal{X}$ of attributes over which policy conditions are defined is a union of data subject attributes (e.g. opt-in/opt-out consents, user location), data accessor attributes (e.g. accessor's role and location), and system attributes (e.g. availability/security zone where the system is located in). The set of actions $\mathbb{A}$ of interest in this paper is the set $\{\texttt{KEEP}, \texttt{MASK}\}$, where the action $\texttt{KEEP}$ allows access to the target data while $\texttt{MASK}$ redacts the data. 
For simplicity, we only consider a specific form of data redaction in this paper, where the target data element is either replaced with $\texttt{NULL}$ value or the tuple containing the target data element is filtered from the result set. 
Note that the same model can be extended to define and support other types of actions like truncating, or partially masking the data.

Without loss of generality, all policies considered in the remainder of the paper are \texttt{KEEP} policies i.e. the data element referenced by a policy label is preserved if the policy condition evaluates to true and masked otherwise.
We further restrict the policy conditions to be based on data subject consents only. 
This restriction is motivated by the fact that data subject consents are the most common basis for data access policies at LinkedIn. 
We therefore focus our attention to those consent-based policies for the remainder of this paper. 

In doing so, we aim to highlight:
\begin{enumerate}
\item the scalability challenges that need to be addressed to enforce consents of a large population of data subjects during data accesses, and 
\item the unique challenges such policies introduce for enabling fine-grained data masking. 
\end{enumerate}
\vspace{-0.5em}
\subsection{Data-Masking Views} \label{subsec:views}
Data Guard uses views as the interface for access control. Each data-masking view is a SQL representation of applicable data policies on a table for a given purpose. The applicable policies for a given purpose are found by matching policy labels assigned to fields in a table against the labels assigned to policies. The matched policies are then translated to SQL by the {\em Policy Compiler}, the implementation of which will be discussed in detail in Section~\ref{subsubsec:pc}. 

We illustrate an example data-masking view created by Data Guard in Figure~\ref{fig:SystemOverview}. 
Let us assume that the table $T_1$ shown in the figure is keyed by an $id$ column containing data subject identifiers. 
The figure also shows that $T_1$ contains a column $col_2$ that has $education$ data with an applicable policy for the $ads$ purpose. 
For simplicity, let us assume that all data subject attributes referenced in policy definitions are stored in the $member\_settings$ table which is also keyed by the data subject identifier. 
A naively implemented policy compiler generates the following view SQL for $ads$ purpose which masks $col_2$ based on the value of the attribute $allowGenderForAds$ in the $member\_settings$ table:
\lstdefinestyle{sqlstyle}{
    language=SQL,
    basicstyle=\ttfamily\footnotesize,
    keywordstyle=\color{blue},
    commentstyle=\color{gray},
    stringstyle=\color{orange},
    showstringspaces=false,
    morekeywords={SELECT, FROM, WHERE, AND, OR, INSERT, INTO, UPDATE, DELETE}, 
    frame=single, 
    breaklines=true, 
}
\lstset{style=sqlstyle}
\begin{figure}[ht] 
\vspace{-1em}
\begin{lstlisting} 
SELECT T1.id, T1.col1, CASE WHEN allowGenderForAds = true THEN T1.col2 ELSE NULL END AS col2
FROM T1 JOIN member_settings T2 ON T1.id = T2.id;
\end{lstlisting}
\vspace{-1em}
 \caption{A data-masking view for $ads$ purpose}
\label{fig:viewsql}
\vspace{-1em}
\end{figure}

The view SQL shown in Figure~\ref{fig:viewsql}, while inefficient, has the important property that it is {\em schema-preserving} i.e. it has the same schema as the underlying table $T_1$. This invariant is maintained across all views created by Data Guard. The schema-preserving property ensures that applications can switch consumption from tables to views without needing to make code changes. Each view generated by Data Guard is registered within a purpose-specific database (e.g. $ads.V_1$ in Figure~\ref{fig:SystemOverview}), which facilitates search and discovery of views in the data catalog. In Section~\ref{sec:impl}, we discuss a number of optimizations which will be used to rewrite the view SQL in Figure~\ref{fig:viewsql} in order to make it performant.

The decision to use SQL views as the interface for access control was motivated by the following reasons: 
\begin{itemize}
\item {\em Portability}:  Views are engine-agnostic and work out of the box with engines like Spark and Trino.
\item {\em Debuggability}: Views make masking logic visible to end users and allow consumers to reason about {\em what} data is being filtered and {\em why}.
\item {\em Version control}: Views can be versioned in the same manner as software artifacts. Thus, changes to view logic (due to policy and label changes) can be tested by data consumers before they are deployed in production. 
\item {\em Agility:} Because the policy compiler is independent of the compute engine, implementing changes there—such as generating optimized data-masking views—is faster and less risky than modifying the compute engine, which must support many other use cases beyond access control.
\item {\em Optimizations:} A dedicated policy compilation layer provides the flexibility of implementing custom optimizations (e.g. bitmap optimization discussed in Section~\ref{subsubsec:bitmap}) when generating view SQL. Such use-case specific optimizations are otherwise non-trivial to implement inside general-purpose compute engines. 
\end{itemize}
As shown in Figure~\ref{fig:SystemOverview}, accesses to tables are routed to an appropriate masking view based on the purpose of access using a component called {\em ViewShift}.  ViewShift along with the schema-preserving property of the masking views allows applications to seamlessly switch consumption from tables to views without incurring significant migration costs. 
\vspace{-0.5em}

\subsection{Field Paths} \label{subsec:fieldpaths}
\subsubsection{Motivation}
Existing solutions~\cite{Stonebraker74, Rizvi04,Xue23, Agrawal05} support data masking at row, column and cell-level granularity. These masking operations are typically accomplished using traditional projection and selection operators. 
As mentioned in Section~\ref{intro}, Data Guard's access control model supports masking data at much finer granularities than is possible with previous solutions. This need for finer grained data masking is motivated by prevalence of non-relational data in LinkedIn's warehouse (and indeed, many modern data warehouses~\cite{Snowflake} and lakehouses~\cite{Databricks}). It is common to model data using relation-valued attributes and {\em collection} types such as arrays and maps.

In many ML applications, user attributes and features are stored in a structured yet sparse format, such as arrays of key-value pairs or nested feature maps, avoiding the need for wide and constantly evolving tables as new features are added. Additionally, some tables combine data from different sources in different rows, or as different elements in arrays. The ability to represent, label and mask data within arrays or maps conditionally provides fine-grained control over compliance while preserving the efficiency and convenience of representation. Masking data selectively inside such data structures cannot be done using relational operators such as projection and selection alone.  

Let us consider an example relation $\mathcal{R}$ with the following schema:
\vspace{-0.6em}
\begin{tabbing}
    \hspace{1cm} \= \hspace{1cm} \= \kill
    \textbf{$col_1$}: \> \small\texttt{VARCHAR} \\
    \textbf{$col_2$}: \> \small\texttt{STRUCT<}$field_{21}$\small\texttt{:BIGINT, }$field_{22}$\small\texttt{:VARCHAR>} \\
    \textbf{$col_3$}: \> \small\texttt{ARRAY<STRUCT<}$field_{31}$\small\texttt{:VARCHAR, }$field_{32}$\small\texttt{:DOUBLE>>} \\
    \textbf{$col_4$}: \> \small\texttt{MAP<VARCHAR, ARRAY<STRUCT<}$field_{41}$\small\texttt{:VARCHAR, } \\
                    \> $field_{42}$\small\texttt{:BOOLEAN>>>}
\end{tabbing}

\definecolor{gray1}{gray}{0.9}
\definecolor{gray2}{gray}{0.8}
\definecolor{gray3}{gray}{0.7}
\definecolor{gray4}{gray}{0.6}
\definecolor{gray5}{gray}{0.5}
\begin{figure}[htbp]
\vspace{-1em}
\resizebox{0.5\textwidth}{!}{
\begin{tabular}{|c|c|c|c|c|c|c|c|}
    \hline
    \multirow{2}{*}{\textbf{$col_1$}} & \multicolumn{2}{c|}{\textbf{$col_2$}} & {\textbf{$col_3$}} & {\textbf{$col_4$}} \\ \cline{2-3}
    & \textbf{$field_{21}$} & \textbf{$field_{22}$} &   &  \\ \hline
    abc \cellcolor{gray1} & 123 & foo & 
    \begin{tabular}{|c|c|}
        \hline
        \textbf{$field_{31}$} & \textbf{$field_{32}$} \\ \hline
        s1 \cellcolor{gray4} & \cellcolor{gray4} 113.2 \\ \hline
    \end{tabular} &  
    NULL \\ \hline
    def \cellcolor{gray2} \cellcolor{gray2} & 243 \cellcolor{gray2}  & bar \cellcolor{gray2}  & NULL \cellcolor{gray2}  & NULL  \cellcolor{gray2} \\ \hline
    ghj \cellcolor{gray1} & 123 \cellcolor{gray3} &  bar \cellcolor{gray3} & 
    \begin{tabular}{|c|c|}
        \hline
        \textbf{$field_{31}$} & \textbf{$field_{32}$} \\ \hline
        s1 \cellcolor{gray4} & 345.2 \cellcolor{gray4} \\ \hline
        s3 & 212.0 \\ \hline
    \end{tabular} & 
    \begin{tabular}{|c|c|c}
        \hline
        \textbf{key} & \textbf{value} \\ \hline
        k1 & \begin{tabular}{|c|c|}
            \hline
            \textbf{$field_{41}$} & \textbf{$field_{42}$} \\ \hline
            v1 & true \\ \hline
            v2 & false \cellcolor{gray5} \\ \hline
        \end{tabular} \\ \hline
        k2 & NULL \\ \hline
    \end{tabular} \\ \hline
\end{tabular}
\vspace{1em}
}

\resizebox{0.3\textwidth}{!}{
\begin{tabular}{|c|l|}
\hline
\cellcolor{gray1} &  $\$.col_1$ \\
\hline
\cellcolor{gray2} & $\$.[?(@.col_1='def')]$ \\
\hline
\cellcolor{gray3} &  $\$.[?(@.col_1='ghj')].col_2$ \\
\hline
\cellcolor{gray4} & $\$.col_3.[item].[?(@.field_{31}='s1')]$  \\
\hline
\end{tabular}
}
\caption{Example of a nested relation with arrays, structs, maps.}
\label{fig:relationexample}
\vspace{-0.8em}
\end{figure}

In order to mask data at a sub-cell granularity (e.g. $field_{31}$, $field_{42}$ shown in Figure~\ref{fig:relationexample}), we need an operator for selecting this data. There have been prior proposals~\cite{Colby90} which introduce recursive selection and projection operators, which can be used to select and mask data at sub-cell granularity. While there have been efforts to extend SQL language to support recursive operations~\cite{sql1999}, very few commercial and open-source systems support such extensions. While introducing these extensions to existing open source systems like Spark and Trino is in theory possible, it would require a significant modification of the SQL standard adopted by these systems and a non-trivial investment of effort to drive alignment across the open-source community. On the other hand, there have been numerous path DSLs such as XMLPath \cite{XPath} and JsonPath \cite{rfc9535} that have been developed and successfully adopted to support element-wise operators. Given these trade-offs, using a DSL to evaluate field paths efficiently during data access was a cost-effective alternative. Leveraging existing DSLs such as XPath and JsonPath was not an option due to several limitations that prevent their usage for structured data handling. XPath is intended exclusively for processing XML documents. JsonPath is schema-unaware and does not distinguish between types such as maps and structs. Nevertheless, we use JsonPath to guide the design of the field path DSL described next. 

\subsubsection{Field Path Expressions}
In this section, we provide a formal definition of {\em field path} that is used to select data at a sub-cell-level for non-atomic data types such as structs, arrays and maps. 
\vspace{-0.5em}
\begin{definition} \label{def:fieldpath}
A \emph{field path} is a sequence of operators $P_1$, $P_2$, \ldots, $P_n$, where each operator $P_i$ operates on the result set of $P_{i-1}$ and has one of the following types:
\begin{enumerate}
\item {\bf Root access operator}: the first operator in the sequence of operators for any given field path, denoted by a special symbol $\$$, serving as an identity operator returning the input relation.
\item {\bf Transform operator}:  A transform operator is immediately preceded by either a root access or another transform operator and is prefixed with a $.$ symbol. There are three types of transform operators:
    \begin{enumerate}
        \item {\bf Dereference operator}: This operator is of the form \texttt{.<name>} and projects a sub-attribute of a composite attribute. 
        \item {\bf Filter operator}: This operator selects a subset of a relation that satisfies a given condition. The operator is of the form \texttt{[?(<condition>)]}, where condition is a SQL predicate string on the input relation.
        \item {\bf Unnest operator}: a cell-wise transform operator that applies all subsequent operators on each cell of inner relations in collection-type attributes such as arrays and maps. There are three forms of unnest operators:
        \begin{enumerate}
            \item $[item]$: an operator that applies to array types and provides access to the array items, 
            \item $[key]$: an operator that applies to map types and provides access to the map keys, and 
            \item $[value]$: an operator that applies to map types and provides access to the map values.
        \end{enumerate}
    \end{enumerate}
\end{enumerate}
\end{definition}
\vspace{-0.5em}

\subsubsection{Examples}
We next provide examples of field path expressions that demonstrate their ability to select data elements from the different attributes of the relation shown in Figure~\ref{fig:relationexample}.
\begin{enumerate} 
\item A dereference operator following a root access operator can be used to select an attribute from a relation. For example, $\$.col_1$ selects $col_1$.  
\item A filter operator following a root access operator is referred to as a {\em row selector} and has the ability to select rows from a relation. For example, the field path $\$.[?(@.col_1 = \text{`def'})]$ selects rows from the relation satisfying $col_1 = \text{`def'}$.
\item A row selector followed by a dereference operator selects cell values of an attribute from the selected rows. For example, the field path $\$.[?(@.col_1 = \text{`ghj'})].col_2$ selects values of $col_2$ from rows satisfying the condition $col_1 = \text{`ghj'}$.
\end{enumerate}


In order to mask data addressed by a field path expression, we need the ability to assign policy labels to them. Thus, the field paths associated with a relation are stored with the schema of the relation in the data catalog. Some field paths are automatically extracted from the schema when a schema is ingested into the data catalog. Additional field paths such as row selectors are added to the catalog by owners of the schema. 

\subsubsection{Data-Masking Operator} \label{subsubsec:masking_operator}
We next describe how the field path expressions are used to define data masking operations on a data element. Given an input relation $\mathcal{R}$, a data access policy $p$, and a field path $f$, the data-masking operator is a relational algebra function denoted by $mask(\mathcal{R}, p, f)$ and defined as follows:
\vspace{-0.5em}
\begin{equation}
\small
    mask(\mathcal{R}, p, f) = 
    \begin{cases}
    \sigma_{\neg\text{pred OR p.c}}(\mathcal{R}), 
    \quad \text{if } f=\$.[?(pred)] \\
    {\tau_{f \rightarrow m(f)} (\sigma_{\neg\text{p.c}}(\mathcal{R})))} 
    \cup \sigma_{\text{p.c}}(\mathcal{R}), 
    \text{otherwise}
    \end{cases}  
\label{eq:enforcement}
\end{equation}

where $\sigma$ is the standard relational algebra select operator. $\tau$ denotes a transformation operator that masks the data element at path $f$ by applying a masking function $m$ and reassembles the transformed attribute (potentially, a nested relation) back into its original structure. 
Equation~\ref{eq:enforcement} shows that when $f$ does not contain a dereference operator, the masking operator reduces to a {\em row-level mask} and removes the matching rows from the result set. 
Otherwise, the transformation operator functions as a {\em column-level mask} applying the function $m$ to values of $f$. The masking function $m$ in Equation~\ref{eq:enforcement} has the following behavior:
\begin{itemize}
    \item If $f$ is an atomic attribute, $m(f)$ sets the value of $f$ to NULL.
    \item If $f$ is an array element, $m(f)$ removes the element from the array.
    \item If $f$ is a map key, $m(f)$ removes the key value pair from the map,
    \item If $f$ is a map value, $m(f)$ sets the map value to NULL.
\end{itemize}

In summary, field paths and data masking operators give us the capability of masking data at a very fine granularity. The exact implementation of the data-masking operator will be discussed in detail in Section~\ref{subsubsec:pc}. 
\vspace{-0.8em}
\section{Implementation} \label{sec:impl}
\vspace{-0.5em}
\subsection{View Creation}
In this section we first present the Policy Compiler (Section~\ref{subsubsec:pc}), which translates access-control policies into data-masking views. 
This section describes several optimizations (Sections~\ref{subsubsec:bitmap} and~\ref{subsubsec:dedup}) that ensured scalability of data-masking views in production. 
While these optimizations are presented as heuristics in this paper, we plan to formalize them into a general optimization framework in future work. 
We conclude by describing the view maintenance system that handles policy and label changes (Section~\ref{subsubsec:maintenance}).

\subsubsection{Policy Compiler} \label{subsubsec:pc}
The process of translating data policies into SQL code is referred to as {\em policy compilation}. 
A policy compiler takes as input a relation $\mathcal{R}$ (table or view) and a list $\mathcal{L}$ of matching pairs (field path, policy) for $\mathcal{R}$. 
The output of the policy compiler is a SQL string representing a data-masking view for $\mathcal{R}$.

A key step in the policy compiler is to convert the enforcement of a data access policy $p$ on a field path $f$ into a relational algebra plan representing the \texttt{mask} operator.
We focus on the implementation of the \texttt{mask} operator to perform column-level masking i.e. the data element to be masked is inside an attribute of $\mathcal{R}$. 
A naive approach to implementing the data-masking operator for a relation-valued attribute is to unnest the attribute, mask the data at a given field path within the attribute and reconstruct the attribute into its original structure. 
However, this approach produces complex SQL queries for relation-valued attributes, which can be both hard to read and debug especially when involving transformations on deeply nested attributes. 
We therefore implement a scalar version of the \texttt{mask} operator with a user-defined function for column-level masking, which is maintained by Data Guard platform team.

Specifically, consider a field path $f$ inside an attribute $\mathcal{A}$ of a relation $\mathcal{R}$ and a boolean condition $cond$, 
the following UDF implements a scalar version of the data masking operter $mask$: 
\textit{\texttt{MASK\_FIELD\_IF(cond: \texttt{BOOLEAN}, $\mathcal{A}$: ANY, $f$: VARCHAR)}} \\
If the condition is true, the UDF performs an in-place update on attribute $\mathcal{A}$ setting the value at path $f$ to \texttt{NULL}, instead of a sequence of unnest, transform, and nest operations. If the condition is false, the UDF returns the original value of $\mathcal{A}$. 
Therefore, the return type of the UDF is the same as the data type of $\mathcal{A}$ to ensure the view schema is identical to the underlying table schema.

The data-masking operator introduced in Equation~\ref{eq:enforcement} can thus be implemented via a simple UDF call in SQL as follows: 
\begin{lstlisting} 
MASK_FIELD_IF(p.c, A, f)
\end{lstlisting}
where A is the root attribute of the field path $f$.

\subsubsection{Consent Bitmaps} \label{subsubsec:bitmap}
Recall from Section~\ref{subsubsec:policies} that the most common type of data access policies in our system are consent-based, where the policy condition involves checking preferences of data subjects stored in a separate table, referred to as the {\em lookup table}. As described in Section~\ref{subsec:views}, a naive implementation of policy compiler can produce views that join a table containing data subject information with the lookup table using the data subject id as the join key. At LinkedIn, this table stores the consents of more than 1 billion members and any joins against this table result in expensive data shuffles. In Spark for instance, such joins are implemented as Sort Merge Joins since the size of both the tables involved in the join operation are too large to be broadcast using Broadcast Hash Joins. Optimizing the view SQL for fetching attribute values from the lookup table is therefore critical to minimize the cost of adopting views.

In order to address this scalability challenge, we pre-compute a consent bitmap for every consent attribute referenced in a policy definition and leverage statistics associated with each consent to minimize the bitmap size. Specifically, we use a "true bitmap" (bitmap of data subject identifiers whose consent value is true) if a minority of users have consent values set to true, and a false bitmap otherwise. Therefore, the number of user ids in each bitmap is at most half of the total number of users. Given that data subject identifiers at LinkedIn are numeric (a scenario we expect to be common), we use the Roaring Bitmap implementation~ \cite{Lemire16} to compute the bitmaps as it achieves superior compression ratios compared to other compression methods. As an example, a Roaring bitmap of 200 million 8-byte integers that represent user ids has a size of around 134MB, which is small enough for the bitmaps to be loaded into memory on the worker nodes in Spark and Trino. In cases where identifiers are non-numeric, we can employ minimal perfect hashing functions~\cite{Belazzougui09} to map keys into consecutive integers and then use Roaring Bitmap on the resulting integers.

These bitmaps are then saved in a storage system like HDFS and loaded at access time for fast in-memory lookup via a UDF. The bitmap implementation is much cheaper compared to a join-based solution, as we only introduce a small memory overhead for storing bitmaps in each worker's memory. In order to abstract away the storage system implementation details, we wrap the bitmap loading and lookup actions into a BitmapManager interface, thus ensuring that the bitmap solution can be adopted across a range of different storage systems.

To ensure freshness of bitmaps, a bitmap computation job is scheduled periodically to reflect the latest user consents in the generated snapshot. The period is decided based on legal requirements. Each snapshot of a bitmap is addressed using a combination of the bitmap name and a timestamp when the bitmap was generated. The different snapshots of a bitmap are stored in a time-partitioned folder. The bitmap lookup is wrapped inside a UDF with the following signature: 
\texttt{HAS\_USER\_CONSENT(consents: \texttt{VARCHAR}, user\_id: \texttt{BIGINT}, access\_time: \texttt{TIMESTAMP})}, 
where:
\begin{itemize}
    \item \emph{consents}: a string to represent a list of consent names, which are used to load corresponding bitmaps into executor memory.
    \item \emph{user\_id}: the identifier column of the table identifying the data subject whose consents need to be looked up, 
    \item \emph{access\_time}: the timestamp for determining the specific snapshot of a bitmap to load. This parameter gives developers the ability to replay a query at a later time (aka {\em time travel}). By default, the data-masking views use the built-in SQL function \texttt{CURRENT\_TIMESTAMP()} for the $access\_time$ parameter. The $access\_time$ parameter ensures that the same snapshot of bitmap is loaded across all the worker nodes for a given query. 
\end{itemize}

\subsubsection{Masking deduplication} \label{subsubsec:dedup}
Given the scale of LinkedIn's data warehouse and the diversity of policies that need to be enforced, it is common to find tables in our warehouse containing attributes that are subject to multiple policies. 
Further, a single policy may apply to multiple attributes within the table. A naive implementation of the policy compiler can result in redundant application of the \textit{\texttt{MASK\_FIELD\_IF()}} UDF. 

There are two scenarios that can result in redundant masking operations on a field path: 
\begin{itemize}
    \item If a nested field needs to be masked based on one or more policy conditions, each of which is applicable to an ancestor, then a masking operation on the nested field is unnecessary.  For example, masking operation on the field path in row 2 of Table~\ref{tab:optimization} is redundant given an identical policy condition applicable to the root path \$. Similarly, the masking on field path in row 4 is redundant given an identical policy condition applicable to its parent in row 3.   
    \item If a field in a table has two or more applicable policies and the set of consents used in policy $p_i$ is a superset of the set of consents used in a different policy $p_j$, then the masking with policy $p_j$ is unnecessary. For example, rows 5 and 6 of Table~\ref{tab:optimization} show two different policies $p_3$ and $p_4$ applied to the same field path with overlapping consents. It is easy to see that policy condition checks in row 2, 4, 6 are redundant and can be eliminated.
\end{itemize}

\begin{table}[h!]
    \centering
    \resizebox{\columnwidth}{!}{
    \begin{tabular}{|c|c|l|c|c|}
        \hline
        \textbf{No} & \textbf{Field Path} & \textbf{Policy condition} & \textbf{Policy} \\ \hline
        1 & $\$$ & $consent_1$ & $p_1$ \\ \hline
        2 & $\$.col_2.field_{21}$ & $consent_1$ & $p_1$ \\ \hline
        3 & $\$.col_3$ & $consent_2$ & $p_2$ \\ \hline
        4 & $\$.col_3.[item].[?(@.field_{31}='s1')]$  & $consent_2$ & $p_2$ \\ \hline
        5 & $\$.col_4.[value]$ & $consent_3$ AND $consent_4$ & $p_3$ \\ \hline
        6 & $\$.col_4.[value]$ & $consent_3$ & $p_4$ \\ \hline
    \end{tabular}
    }
    \caption{Maskings for different field paths with policies}
    \label{tab:optimization}
\end{table}

\begin{figure}[ht]  
    \begin{subfigure}[b]{0.58\columnwidth}
     \includegraphics[width=\linewidth]{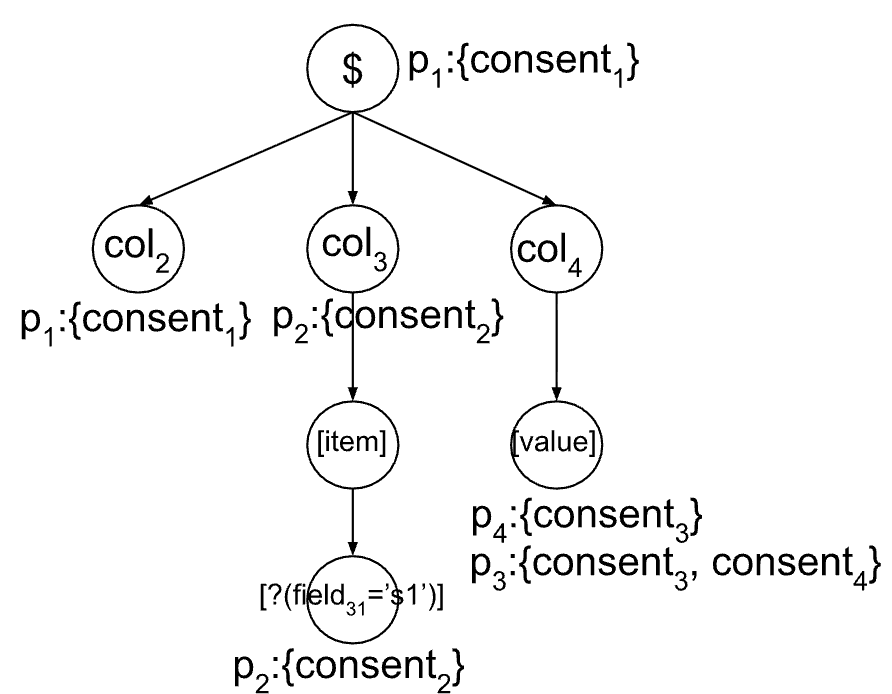}
     \caption{Before pruning}  
     \label{fig:schematree}  
     \end{subfigure}
     \hfill 
     \begin{subfigure}[b]{0.38\columnwidth}
     \includegraphics[width=\linewidth]{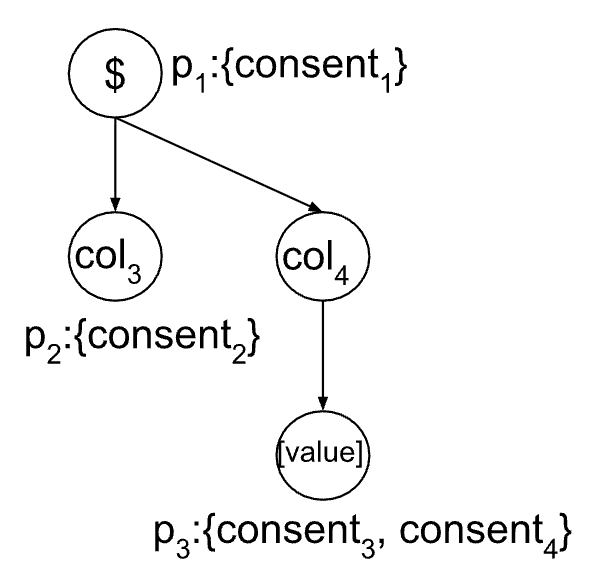}
     \caption{After pruning}  
     \label{fig:prunedschematree}  
     \end{subfigure}
     \vspace{1ex}
     \caption{A schema tree constructed from field paths in Table~\ref{tab:optimization}.}
     \label{fig:pruningtrees}  
\end{figure}

To eliminate such redundant masking operations, we construct a schema tree from a given list of field paths and their applicable policies. 
A schema tree is one where each node represents an operator of a field path expression defined in Definition~\ref{def:fieldpath} and nodes corresponding to successive operators of a field path expression have a parent-child relationship between them. 
A node that is a final operator of a field path has an attribute to store the list of all policies applicable to that field path.
Figure~\ref{fig:schematree} shows an example schema tree constructed from field paths in Table~\ref{tab:optimization}. 

After constructing the schema tree, we perform a pre-order traversal to prune redundant policies as detailed in Algorithm~\ref{alg:policy_pruning}. The algorithm computes the minimal set of policies required to cover the set of consents of a node that do not overlap with any of its ancestors. This problem can be mapped to the well-known minimum set cover problem, which is an NP-complete problem~\cite{Garey90}. We therefore use a greedy approximation algorithm that chooses policies with largest number of non-overlapping consents first.
Figure~\ref{fig:prunedschematree} demonstrates the schema tree produced by applying Algorithm~\ref{alg:policy_pruning}. 
Maskings in rows 1, 3, and 5 are chosen to be retained by the algorithm. 

\begin{algorithm}
    \caption{Prune redundant policies}
    \label{alg:policy_pruning}
    \begin{algorithmic}
        \STATE \textbf{Input:} A schema tree with root $r$.
        \STATE \textbf{Output:} A new schema tree with redundant policies pruned.
        \STATE \textbf{function} \textsc{PrunePolicies}($n$, $C$)
        \STATE \hspace{1em} $P \gets n.policies$ \hfill \text{//Get node $n$'s policy set}
        \STATE \hspace{1em} $C_n \gets \emptyset$ \hfill \text{//Initialize node $n$'s consent set}
        \STATE \hspace{1em} \textbf{for} $p_i \in P$ \textbf{do}
        \STATE \hspace{2em} $C \gets C_n \cup consents(p_i)$ 
        \STATE \hspace{1em} \textbf{end for}
        \STATE \hspace{1em} $\Delta \gets C_n \setminus C$ \hfill \text{//consents non-overlapping with $n$'s ancestors}
        \STATE \hspace{1em} $R \gets \emptyset$ \hfill \text{// Initialize the retained policy set}
        \STATE \hspace{1em} \textbf{while} $\Delta \neq \emptyset$
        \STATE \hspace{2em} select $p \in P$ that maximizes $|consents(p) \cap \Delta|$
        \STATE \hspace{2em} $\Delta \gets \Delta \setminus consents(p)$
        \STATE \hspace{2em} $R \gets R \cup \lbrace p \rbrace$
        \STATE \hspace{2em} $P \gets P \setminus \lbrace p \rbrace$
        \STATE \hspace{1em} \textbf{end while}
        \STATE \hspace{1em} $n.policies \gets R$ \hfill \text{//Update node $n$'s policy set}
        \STATE \hspace{1em} \textbf{for} $c \in n.children$ \textbf{do}
        \STATE \hspace{2em} \textsc{PrunePolicies}($c$, $C \cup C_n$)
        \STATE \hspace{1em} \textbf{end for}
        \STATE \textbf{end function} 
        \STATE \textsc{PrunePolicies}($r$, $\emptyset$)
    \end{algorithmic}
\end{algorithm}

Suppose $col_1$ of $\mathcal{R}$ is labeled as a \emph{user id} column, the resulting data-masking view SQL produced by the policy compiler is as follows:
\begin{lstlisting} 
SELECT col1, col2, 
 MASK_FIELD_IF(HAS_USER_CONSENT('consent2', col1, CURRENT_TIMESTAMP()), col3, '$.col3') AS col3,
 MASK_FIELD_IF(HAS_USER_CONSENT('consent3_consent4', col1, CURRENT_TIMESTAMP()), col4, '$.col4.[value]') AS col4
FROM R
WHERE HAS_USER_CONSENT('consent1', col1, CURRENT_TIMESTAMP());
\end{lstlisting}

\subsubsection{View Maintenance} \label{subsubsec:maintenance}
We implement a system for creating and updating of data-masking views across tables in our warehouse. At a high-level, this system performs the following steps: 
\begin{enumerate} 
    \item Generates a candidate list of warehouse tables along with a map of field paths to matching policies for each candidate table. This is accomplished by joining: (i) an inventory of warehouse tables, (ii) a metadata table that contains information about policy labels assigned to each field path of a table, and (iii) a table containing policy labels and policy definitions. 
    \item For each candidate table $T$ computed in the previous step and each candidate purpose $P$:
    \begin{enumerate}
    \item the system checks if it needs to create or update a masking view for $T$. Specifically, view update for $T$ is triggered due to: 
    \begin{itemize}
        \item schema changes to the underlying tables (e.g. new column additions), 
        \item changes to policy labels assigned to field paths, or 
        \item new policy additions or changes to existing policy definitions for purpose $P$.
    \end{itemize}
    \item If a view update is deemed necessary, the system invokes the policy compiler to generate a new view definition, and 
    \item registers the updated view definition with the view catalog (a Hive-based~\cite{hive} catalog in our ecosystem). 
    \end{enumerate}
\end{enumerate}

To ensure smooth rollout of views and enable rollbacks in case of errors, each data-masking view is versioned according to the semantic versioning scheme~\cite{semver}. 
Data pipelines are routed to the latest view versions by default. 
We also provide developers with the option to pin their pipelines to specific versions of views and be notified when newer versions of views become available. 
In this mode, developers can test their pipelines against the new view versions before deploying the change to production.
Older versions of views are garbage collected periodically to minimize resource consumption due to stale views. 

\subsection{View Consumption}
At LinkedIn, developers write data processing applications using Apache Spark and Trino as the underlying compute engines. To ensure ease-of-use of data-masking views in applications, we implemented two important optimizations: (i) dynamic table-to-view routing, and (ii) multi-engine support, each of which will be discussed in detail in this section.

\subsubsection{Dynamic Table-to-View Routing}
A key design choice of our system was to ensure access control policies are automatically enforced during data accesses by redirecting table accesses to the appropriate data-masking views determined by the purpose of access. 
This minimizes overhead for existing developers to ensure their accesses remain compliant. 
Recall from Section~\ref{sec:overview} that the table to view routing at access time is accomplished using a component called {\em ViewShift}. 
At LinkedIn, we implement this component as a plugin inside existing engine catalog implementations. 
ViewShift intercepts \texttt{loadTable()} API calls to the catalog from applications and returns the corresponding data-masking view instead. The ViewShift plugin exposes the following API: \\
{\small \texttt{ViewIdentifier getView(TableIdentifier tableName, Map\textless{}String, String\textgreater{} contextMap)}}\ \\
where \texttt{contextMap} parameter is query context map that contains attributes such as the purpose of the access (obtained from the identity of the developer that submits the query), the optional pinned version of the data masking view, and the environment where the application is running. 

Every invocation of the \texttt{getView()} API is published to a logging system as a log entry containing the contextual information inside the \texttt{contextMap}. These logs are consumed by downstream monitoring applications that monitor data accesses across the warehouse. 

\subsubsection{Multi-engine support}
Data Guard effectively supports diverse data processing engines (Spark, Trino) with differing SQL dialects and UDF systems - a common challenge in large organizations. To ensure consistent enforcement across these systems without duplication, we leverage two open-source technologies built in-house:

\noindent\textbf{Coral} \cite{coral}, our SQL translation tool, converts data-masking views from a centralized Hive SQL dialect to engine-specific dialects at access time using Apache Calcite~\cite{calcite}. It parses SQL from source dialect into a relational algebra representation, applies transformations to address variations in SQL dialects, and converts the representation into the target SQL dialect.

\noindent\textbf{Transport UDF} \cite{transport} framework enables writing core UDF logic once, and making it reusable across multiple engines through light wrappers. By abstracting engine-specific implementations behind common interfaces, Transport eliminates redundant code while maintaining native performance by directly accessing each engine's type system without data conversions.

\section{Performance Evaluation}
When data-masking views are used to enforce access control policies, it is important to ensure that the performance overhead incurred during data access is minimal.
In this section, we first study the performance overhead of data-masking views through a micro-benchmark to understand how the view performance is impacted given the access control policies being applied.
We then share some statistics on the usage of data-masking views in our production warehouses at LinkedIn.
\subsection{Micro-benchmark}
Since there has been no prior standard benchmark to evaluate the performance of access control policies, we develop our own micro-benchmark to 
understand the performance characteristics of the data-masking operation implemented using the \textit{\texttt{MASK\_FIELD\_IF()}} UDF. 
The primary goal of this micro-benchmark is to isolate and measure the performance overhead introduced by the data-masking view compared to the baseline of
accessing the original table directly without any masking. In the micro-benchmark, we ignore post data-access operations such as joins and aggregations, 
as these operations are application-specific and orthogonal to the data-masking operation.

We developed a synthetic data generation tool to produce datasets with a variety of schemas and data types commonly used in practice, including:
\begin{itemize}
    \item {\bf Primitive Fields:} columns of string type to simulate simple flat data structures. 
    \item {\bf Flat Structs:} structs with varying number of primitive fields.
    \item {\bf Nested Structs:} structs with nested fields of varying depths.
    \item {\bf Arrays:} columns of array type, with varying array sizes. Each array element is a struct with multiple primitive fields.
\end{itemize}

For our benchmarks, we generate a synthetic dataset with 10M rows, stored in ORC format with ZLIB compression — the same format used in LinkedIn's data warehouse. 
While the reported results are based on ORC, we expect similar trends for other widely used formats such as Parquet.

In each of the following experiments, we measure the overhead of data-masking views by running two types of queries on the generated dataset:
\begin{enumerate}
    \item \textbf{Baseline Query:} Reads the original table directly, applying an "identity" UDF to the target field(s). This UDF simply returns the input value unchanged, serving as a control to measure the cost of data access without masking.
    \item \textbf{Masking Query:} Reads the same table but with data access policies applied on the targeted fields, resulting in a data-masking view with \texttt{MASK\_FIELD\_IF()} UDF calls on those field(s).
\end{enumerate}
The overhead of data masking is computed as the relative difference in CPU time per row between the Masking Query and the Baseline Query.
All experiments are run on a Spark cluster with 20 worker nodes, each with 16 vCPUs and 64 GB memory. Each experiment is repeated 5 times and the average CPU time is reported.

\subsubsection{Impact of Data Type} 
\begin{figure}[ht]  
    \centering
    \includegraphics[width=0.4\textwidth]{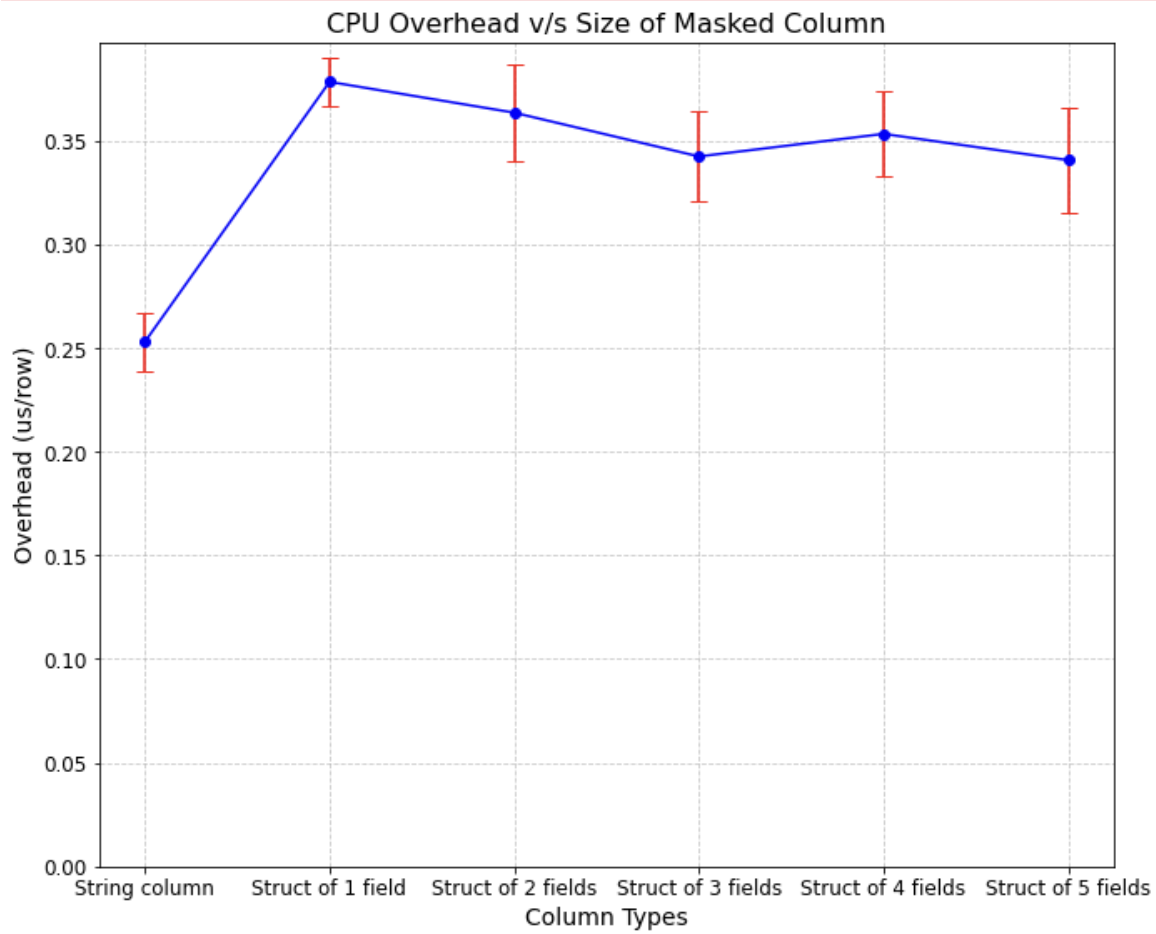}
    \vspace{-0.4em}  
    \caption{Impact of field size on CPU overhead}
    \label{fig:PerfRedactedFieldSize}
    \vspace{-1em}  
\end{figure}

We measure the overhead of applying a data access policy on a column as a function of its size. In
Figure~\ref{fig:PerfRedactedFieldSize}, we show the CPU overhead of masking a flat struct column containing varying number of fields. 

Figure~\ref{fig:PerfRedactedFieldSize} shows that masking a primitive field (such as a string) incurs slightly less overhead compared to masking complex fields like structs. 
Additionally, increasing the number of fields in the column does not significantly affect the overhead.

\subsubsection{Impact of Field Depth} 
We measure the overhead of applying a data access policy on a field as a function of its depth within a nested struct.
\begin{figure}[ht]  
    \centering
    \includegraphics[width=0.4\textwidth]{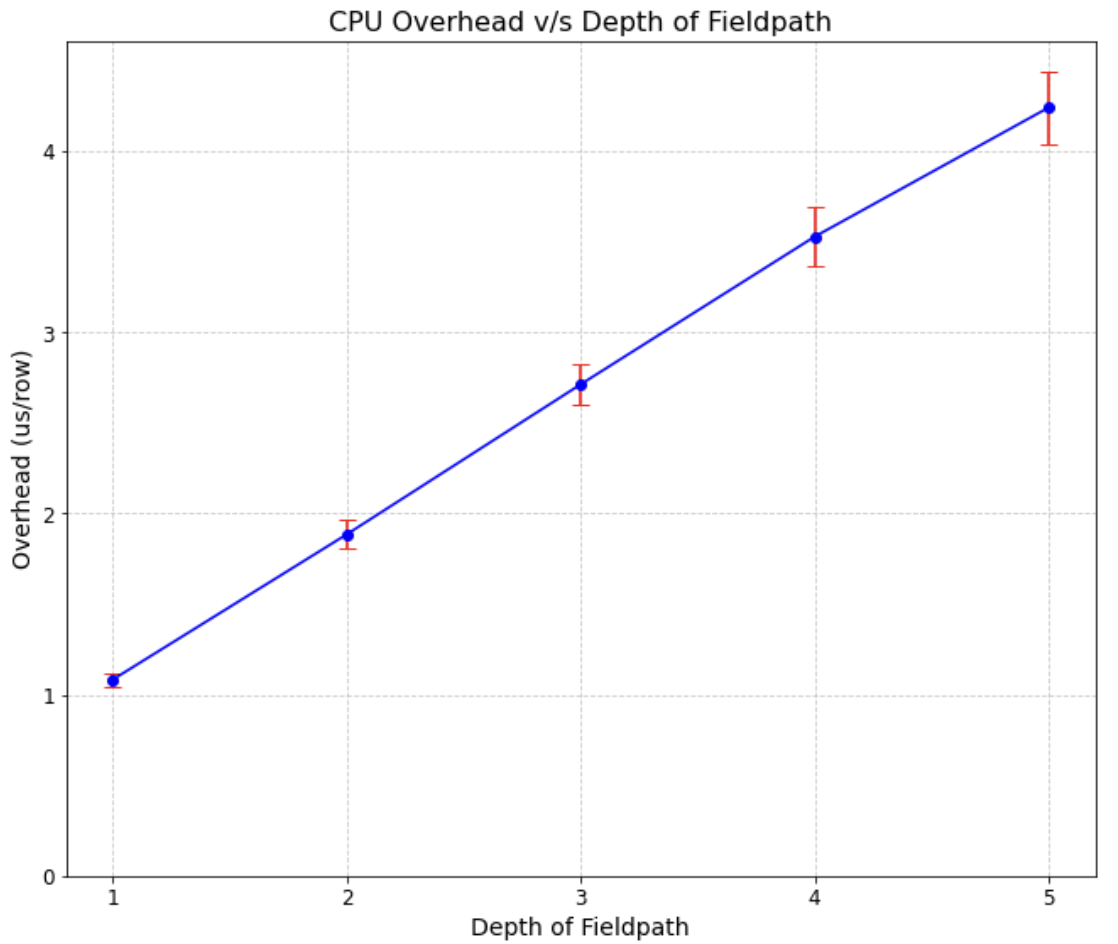}
    \vspace{-0.4em}  
    \caption{Impact of field path depth on CPU overhead}
    \label{fig:PerfVaryingDepthsResult}
    \vspace{-1em}  
\end{figure}

Figure~\ref{fig:PerfVaryingDepthsResult} demonstrates that the overhead increases linearly with the depth of the enforced fields. 
This is because struct objects are organized hierarchically, requiring the \texttt{MASK\_FIELD\_IF()} UDF to traverse the tree structure to locate and mask the target field. 
After masking, each parent struct along the path must be reconstructed to reflect the change, as objects passed to the UDF are immutable.

\subsubsection{Impact of number of policies applied} 
In order to understand the performance of data-masking views as a function of number of policies, we vary the number of invocations of the \textit{\texttt{MASK\_FIELD\_IF()}} UDF on a struct column, one invocation per field of the struct column. In other words, we assume each policy results in an invocation of the masking UDF. 
\begin{figure}[ht]  
    \centering
    \includegraphics[width=0.4\textwidth]{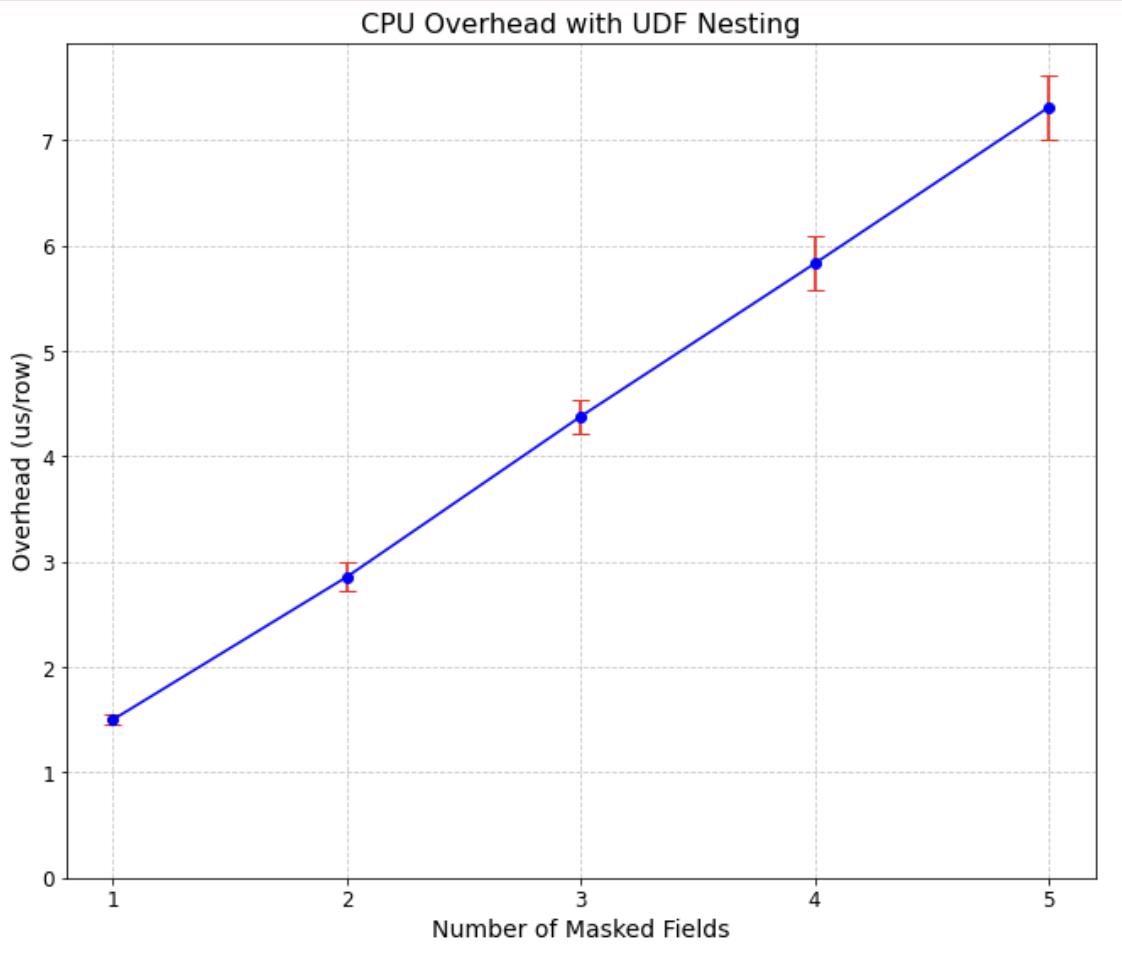}
    \vspace{-0.5em}  
    \caption{Impact of number of applied policies on CPU overhead}
    \label{fig:PerfVaryingRedactedFieldsResult}
    \vspace{-1em}  
\end{figure}

Figure~\ref{fig:PerfVaryingRedactedFieldsResult} shows that the CPU overhead increases linearly with the number of policies applied. 
This is because each invocation of the \texttt{MASK\_FIELD\_IF()} UDF introduces a fixed cost for schema traversal and object transformation. 
Notably, there is potential for optimization when multiple fields within the same struct are masked: batching these operations into a single UDF call can reduce redundant overhead. 
We plan to generalize this optimization in our UDF implementation to leverage the policy compiler's batching capabilities.

\subsubsection{Impact of consent rates} 
In this experiment, we measure the overhead of applying data access policies as a function of the fraction of rows that require masking.  
This fraction depends on whether data subjects consent to allow access to those fields, which we refer to as the {\em consent rate}. 
\begin{figure}[ht]  
    \centering  
    \includegraphics[width=0.4\textwidth]{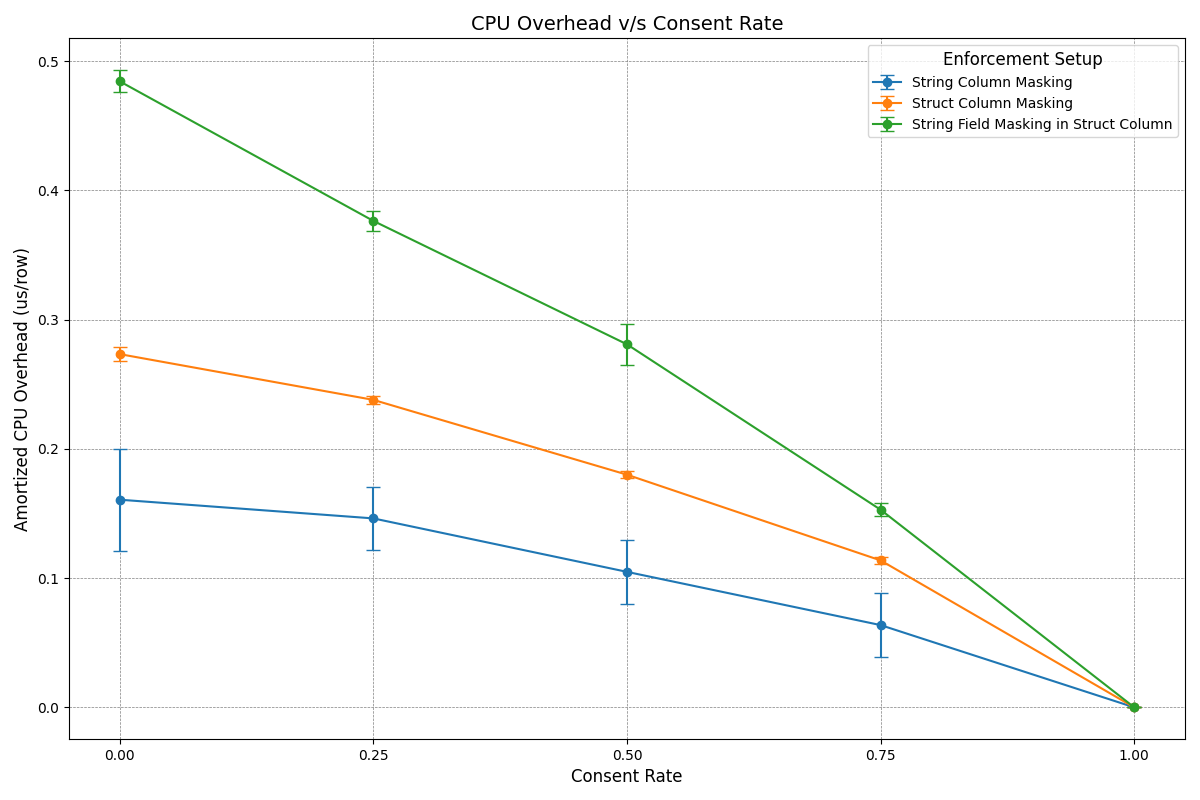}
    \vspace{-1em}
    \caption{Impact of consent rates on CPU overhead}  
    \vspace{-1em}
    \label{fig:PerfVaryingRedactionRate}  
\end{figure}
Figure~\ref{fig:PerfVaryingRedactionRate} contains the micro-benchmark results showing the CPU overhead for different data types as a function of the
consent rate. A consent rate of 1.0 reduces the masking query to the baseline query. As expected, the average CPU overhead decreases
as the consent rate increases for each data type.

\subsubsection{Data Masking in Arrays} 
We next evaluate the overhead of applying data access policies on array-typed columns where each element is of type struct. 
We consider two settings:
\begin{itemize}
    \item \emph{Unconditional masking:} mask a field in each struct for all elements in the array.
    \item \emph{Conditional masking:} mask a field in each struct for elements in the array that satisfy a given condition. 
    This results in an extra overhead of evaluating the condition for each element in the array.
\end{itemize}

\begin{figure}[ht]  
    \centering  
    \includegraphics[width=0.4\textwidth]{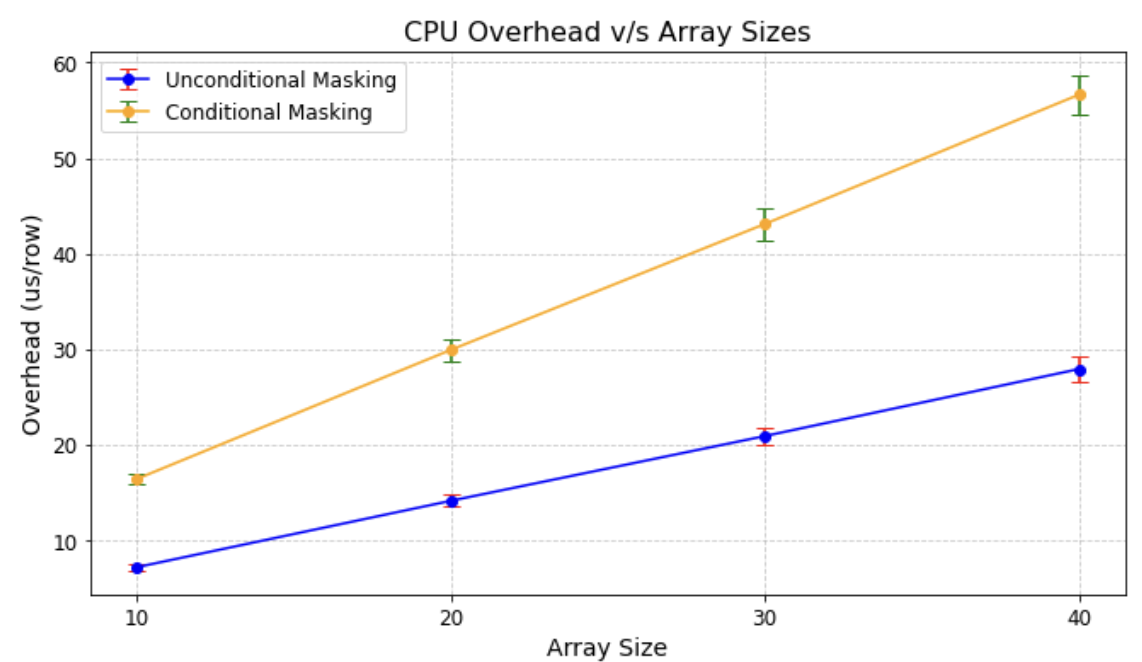}
    \vspace{-0.5em}  
    \caption{Impact of array sizes on CPU overhead}
    \label{fig:PerArrayPrimSec}  
    \vspace{-0.8em}  
\end{figure}

Figure~\ref{fig:PerArrayPrimSec} shows that the masking overhead grows linearly with array size for each field path.
For a given array size, the overhead of unconditional masking is smaller than
conditional masking due to the additional cost of comparing element values.

\subsection{Data Guard in Production}
Data Guard has been deployed as the default access control layer to serve data accesses to LinkedIn's exabyte-scale data warehouse via Spark and Trino, 
ensuring compliance with our data usage policies and regulations. Table~\ref{tab:prod-stats} shows some statistics on the usage of data-masking views 
in our production warehouses.

Based on case studies of some critical usecases, the average CPU overhead of data-masking views per application is 13.2\%, which we consider acceptable given the importance of access control. 
Some data-masking views incur higher overhead when policies target multiple fields or deeply nested structures, consistent with our micro-benchmark results. 
By contrast, many views apply only row-level masking or mask a small number of top-level fields and therefore have much lower overhead.
In some cases, row-level masking can even reduce query cost because filtering out rows decreases downstream work.

\begin{table}[ht]
\centering
\begin{tabular}{|l|r|}
\hline
\textbf{Metric} & \textbf{Value} \\         \hline
Number of purposes supported & 50+ \\ \hline
Number of unique policies enforced & 110+ \\ \hline
Number of tables protected with Data Guard policies & 4,000+ \\ \hline
Number of active data-masking views & 13,000+ \\ \hline
Number of protected daily table accesses  & 20,000+ \\ \hline
Number of protected Spark applications & 2,700+ \\ \hline
Number of protected Trino applications & 800+ \\ \hline
Average number of policies applied per view & 1.7 \\ \hline
Maximum number of policies applied per view & 27 \\ \hline
Average depth of enforced fields & 1.94 \\ \hline
Maximum depth of enforced fields & 9 \\ \hline
Average CPU overhead of data-masking views per application & 13.2\% \\ \hline
\end{tabular}
\caption{Data Guard Production Statistics at LinkedIn}
\label{tab:prod-stats}
 \vspace{-2em}
\end{table}

\section{Practical Considerations} \label{sec:considerations}
In this section, we discuss several practical issues considered while deploying Data Guard at LinkedIn's scale:

\noindent\textbf{Enforcement Verification:} 
Beyond access-time enforcement, Data Guard implements automated verification through continuous monitoring of storage access logs and query execution logs. This detects potential bypasses or accesses to stale views, providing comprehensive audit trails. Our automated remediation system notifies consumers of compliance issues and suggests corrective actions, closing the enforcement loop.

\noindent\textbf{Policy Label Management:} 
At LinkedIn, policy labels are carefully curated and organized into a hierarchical structure referred to as {\em ontology}. An ontology-based organization not only classifies data in our ecosystem, but also captures relationships between these data classes. Ontology also supports advanced use cases at LinkedIn related to reasoning and inference on compliance metadata. We note however that the access control model proposed in this paper does not require an ontology-based organization of policy labels. 

\noindent\textbf{Policy Management:} 
At LinkedIn, purposes are organized into a directed acyclic graph (DAG), where edges denote parent–child relationships. A child may have multiple parents and inherits all parent policies, with conflicts resolved in favor of the more restrictive policy i.e. $MASK$ policies taking precedence over $KEEP$ policies. This hierarchy with inheritance reduces policy duplication across related purposes. Policies are managed like source code: they are versioned, code-reviewed, and synced to an online {\em Policy Catalog} that offers APIs for querying by purpose or label. Policies in Data Guard are intentionally simple, enabling software engineers and product managers—typically feature owners—to author them without requiring policy expertise. 

\noindent\textbf{Developer Experience:} 
Data Guard fundamentally transforms how product teams interact with data compliance requirements. By abstracting compliance logic into centralized policies, teams no longer need specialized privacy knowledge or maintenance overhead of custom application-level implementations. This shift reduced development time from weeks to 2-3 days, as engineers can now focus on product features rather than interpreting complex compliance obligations. Our enhanced data catalog provides interactive masking impact visualizations and clear documentation of view behaviors. 

\noindent\textbf{Engine Optimizations:} 
Implementing masking via UDFs has a limitation that it prevents query optimizations like predicate pushdown or nested column pruning. Our focus so far has been on the APIs for enforcing policies and ensuring a friction-free developer experience. While current performance overhead remains acceptable for production workloads with optimizations implemented so far, we recognize opportunities for further optimizations.

\section{Conclusions And Future Work}
We have presented Data Guard - a fine-grained access control system for large data warehouses. Data Guard's policy language supports specification of access-control rules based on semantic labels used to describe data present in the warehouse tables. Data Guard's policy compiler translates these policies into SQL views which encode the data-masking logic. Further, Data Guard transparently routes table accesses to the appropriate data-masking views based on the accessor's purpose, thus supporting purpose-based access control. Data Guard has been deployed at-scale in LinkedIn and is the primary policy enforcement mechanism for accesses to warehouse tables. We plan to continue work on a number of opportunities for performance optimizations in Data Guard as highlighted in the paper. We also plan to enhance Data Guard capabilities to support privacy-enhancing technologies such as anonymization and differential privacy, and encryption for data storage. Finally, defining a theoretical framework for the fine-grained access control problem and establishing the properties of Data Guard in the context of this framework is an important future direction.

\section{Acknowledgment}
We would like to thank Kapil Surlaker and Raghu Hiremagalur for providing numerous critical inputs on the design of Data Guard. We would like to thank Yong Li, Manasa Subramanian, Divya Singh, Yanwen Lin, Justin Heaton, Vishwaa Patel, Bryan Ji, Qiang Fu, Steve Cao, Wenning Ding, and Carleen Li for their significant contributions to the development of Data Guard. 
Finally, we would like to thank Maneesh Varshney and Chris Harris for their helpful feedback in improving the quality of the paper.

\bibliographystyle{IEEEtran}
\bibliography{main}

\end{document}